\documentclass[final,3p,times,twocolumn]{elsarticle}

\usepackage{graphicx}

\usepackage{amssymb}

\usepackage{subfigure}

\usepackage{lineno}

\journal{Elsevier}

\begin{document}

\begin{frontmatter}



\title{Acoustic transient event reconstruction
and sensitivity studies with the
South Pole Acoustic Test Setup}


\author[jens]{Jens~Berdermann}
\ead{jens.berdermann@desy.de}
\author[icecube]{for the IceCube collaboration}
\address[jens]{DESY, D-15735 Zeuthen, Germany}
\address[icecube]{http://icecube.wisc.edu/}

\begin{abstract}
The South Pole Acoustic Test Setup (SPATS) consists of four strings
 instrumented with seven acoustic sensors and transmitters each, which are
 deployed in the upper 500 m of the IceCube holes. 
Since end of August 2008 SPATS is operating in transient mode, where three
sensor channels of each string, located at three different depth levels, are
 used for triggered data taking within the 10 to 100 kHz frequency
 range.
This allows to reconstruct the position of the source of acoustic signals in the
 antarctic ice with high precision. 
Acoustic signals from re-freezing IceCube holes are identified.
All detected acoustic events seen are associated to sources caused by human
 activities at the South Pole \cite{TRANSIENTS}.
Further, the sensitive volume for $\nu$ interactions outside the IceCube
 instrumented area has been determined by simulation and a flux limit for
 high energy neutrinos was derived.
\end{abstract}


\end{frontmatter}

\linenumbers


\section{Vertex reconstruction}\label{reco}

The reconstruction of acoustic transient events is based on the solution of the
 idealised global positioning equation system
 $$
   (x_n-x_0)^2+(y_n-y_0)^2+(z_n-z_0)^2=[v_s(t_n-t_0)]^2, 
 $$
where four sensor positions and the signal arrival times
 $t_n,x_n,y_n,z_n$ are used in a single reconstruction $n=1,..,4$.
The event vertex is located at the space time point $t_0,x_0,y_0,z_0$.
As an idealisation a sound propagation in ice without refraction and a constant
 velocity of $v_s = 3850~{\rm m/s}$ is used.
The assumption of a constant speed of sound is only suitable for
 events below a depth of around 200 m and leads to a spread of reconstructed
 events for shallower depths, as one can see from simulation.   
Solving the equation system above leads to two exact real solutions, where one
 of them is the event vertex located at the space time point $t_0,x_0,y_0,z_0$
 and the other turns out to be unphysical. 
\begin{figure}[hbtp]
  \centering
    \includegraphics[width=8cm]{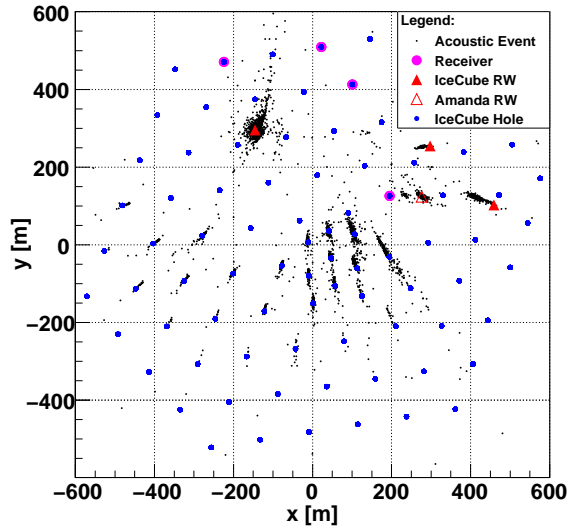}
    \label{fig:all-4s-XY}
  \caption{Horizontal distribution of all transient event vertices recorded
since August 2008. The sources of transient noise are the Rodriguez Wells (RW),
 large caverns melted in the ice for water storage during IceCube drilling, and
 the re-freezing IceCube holes.}
\end{figure}
The recently used SPATS 12-channel configuration \cite{Karg} allows statistical
 predictions by use of all possible sensor combinations per acoustic event.
In case of a noise hit in a sensor the reconstruction algorithm for
 this combination does not converge or the result lies far outside the
 sensitive SPATS area. 
To improve the reconstruction result a cut at the tails of the vertex
 distribution of all sensor combinations has been applied. 
A horizontal distribution of all reconstructed transient events recorded since
 August 2008 is shown in Fig. \ref{fig:all-4s-XY}.  

\section{Events from IceCube holes}\label{holes}
Acoustic events were observed from nearly all IceCube holes drilled when
 transient data taking was active.
The results for different holes are similar, therefore hole 81 is chosen
 below as an example.
\begin{figure}[hbtp]
   \centering   
   \includegraphics[width=6.5cm]{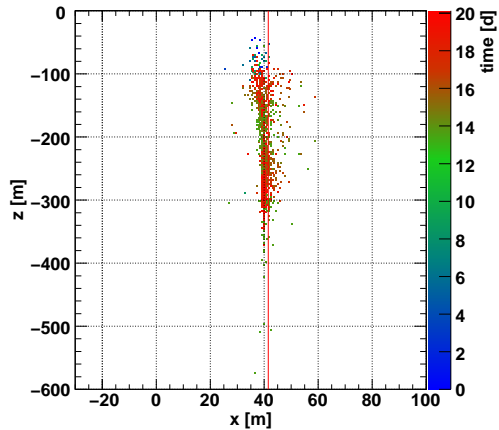}
   \label{fig:h81_xz_color}
   \caption{Events around the region of hole 81 between 10.12 till 31.12.2009
     in the x-z plane. 
     Visible are events before (firn drill), during and after (re-freezing)
 enhanced hot water drilling starts at the 20.12. The red line indicates the
 bore hole position.}
 \end{figure}
Events are observed within 20 days in the hole region ($\pm 20$ m with respect
 to the center of hole 81) during the periods of firn ice drilling
 ($< 50$ m depth), bulk ice drilling ($50-2500$ m depth) and re-freezing.
Fig.~\ref{fig:h81_xz_color} shows the depth position of the events versus
 the x-position, in chronological order. 
A few early events observed at 40 m - 100 m depth are connected
 with noise from the firn drill hole. 
During bulk ice drilling events are found in the same region but also
 at larger depth. 
Strong sound production starts about three days after drilling is
 finished, due to the refreezing process. 

About 30 \% of the registered events from this hole are 
 concentrated in two spots at 120 m and 250 m depth but reaching down to about
 600 m . 
The reason is that the hole doesn't re-freeze homogeneously but forms
 frozen ice plugs between still water filled regions. 
The pressure produced in this way may give rise to cracks near the ice water
 boundary which appear with sound in the 10-100 kHz frequency region. 
Relaxation later continues within ``arms'' freezing upwards to the hole
 surface and down to the lower ice plug.\\
 Besides providing information about the re-freezing process of water-filled 
 IceCube holes one can use the corresponding acoustic events also to understand
 the precision of the vertex localization algorithm. 
The average values for the (x,y) position of hole 81 are determined to:
 40.1$\pm$0.1 m in x and 39.4$\pm$0.1 m in y. 
The width of the distributions is 2.4 m and 4.6 m respectively to be compared
 with a hole diameter of about 70 cm. 
The calculated values deviate from the actual ones by 1.4 m (in x) and 3.9 m (in y). 
The possible reason for this deviation will be discussed in the simulation
 section \ref{sim} below.

\section{Noise from Rodriguez-wells}\label{Rodwells}

When the first acoustic events had been reconstructed during the period from
 August to November 2008 (quiet period) a strong clustering in a certain region
 of the x-y plane at about (-150 m, 300 m) became visible. 
It was found that this was the position of the 2007/08 Rodriguez well (Rod-well
 or RW for short) used for the hot water drilling system.

This type of well has been introduced by Rodriguez and others in the early
 1960s \cite{RODWELL} for water supply at a glacier in Greenland.  
Hot water cycled by a pump system is used to melt ice below the firn layer at
 60-80 m depth to maintain a fresh water reservoir. An expanding cavity is
 formed with a diameter as large as 15-20 m. 
For IceCube and its predecessor AMANDA this technique has been used in
 connection with drilling at the South Pole since mid 1990s.
If the well is used a second time a year later, a second cavern is formed at
 a deeper level. 

Having identified acoustic events arising from the 2007/08 Rod-well, three 
 other event clusters were found, two of them could be attributed to other
 IceCube Rod-wells, from 2006/07 and 2004/05-2005/06. 
The fourth event cluster turned out to be located at the probable position of
 the last AMANDA Rod-well used in the final two drilling seasons up to 2001.
No specific coordinates could, however, be found documented for that position
 anymore. 
The acoustic events from the Rod-wells used during two seasons are located
 at larger depths than those from Rod-wells used only once. 
The latter were seen to emit acoustic signals from smaller and smaller regions
 around the well core and finally stopped, the older one in October 2008, 
the younger one in May 2009. 

\begin{figure}[hbtp]
   \centering
     \includegraphics[width=6.5cm]{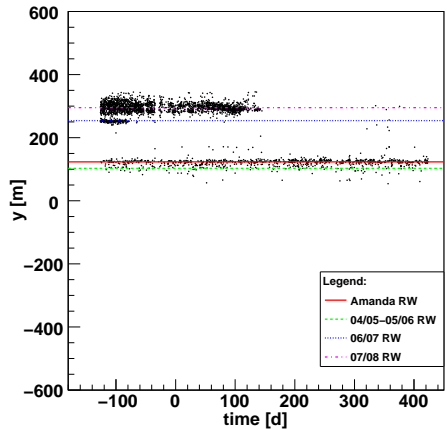}
     \label{fig:RW-yt}
   \caption{Acoustic events distribution for y-coordinate versus time. Lines:
 fitted positions of AMANDA-RW (red), 04/05-05/06 IC-RW (green),
     06/07 IC-RW (blue), 07/08 IC-RW (magenta). 
The zero at the time axis indicates Jan 1, 2009.}
 \end{figure}

In contrast to that, acoustic events are observed until today from the six and
 ten years old deeper wells (see  Fig.~\ref{fig:RW-yt}). 
The mechanism of sound production in and around the Rod-well caverns is still
 under debate in particular for the older wells.  
 \begin{table}[hbtp]
   \caption{\small Parameters of acoustic events from different Rod-wells.
     \label{RW-events}}
   ~\\{\scriptsize        
     \begin{tabular}{|l|c|c|c|c|l|} \hline
       Name & $x_{fit} [m]$ & $y_{fit} [m]$ &
       $z_{fit} [m] $ & used & until \\ \hline \hline
       Amanda & $276.2\pm0.4$ & $123.6\pm1.0$ & $147.3\pm1.1$ &
       2y & now \\ \hline
       IC 04-06 & $412.6\pm5.3$ & $124.0\pm2.4$ &
      $147.8 \pm 11.0$ & 2y & now \\ \hline
       IC 06/07 & $279.6\pm0.4$ & $252.2\pm1.0$ &
       $114.2 \pm 0.7$ & 1y & Oct'08 \\\hline
       IC 07/08 & $-138.6\pm0.4$ & $297.7\pm0.6$ &
       $118.3 \pm 1.0$ & 1y & May'09 \\\hline
     \end{tabular}
    }
 \end{table}

\section{Acoustic event simulation}\label{sim}

A simple approach to acoustic transient event simulation can be done by
 calculating the signal propagation times for the distances
 $d_n=\sqrt{(x_n-x)^2+(y_n-y)^2+(z_n-z)^2}$ between source (e.g. IceCube hole
 at x,y,z) and sensors $n=1,...,n_{max}$ with $\Delta t_n=d_n/v_s$.
The signal is randomly transmitted from a certain cylindrical
 volume (radius 2 m, depth 2000 m) around the source. 
\begin{figure}[hbtp]
   \centering
     \includegraphics[width=8cm]{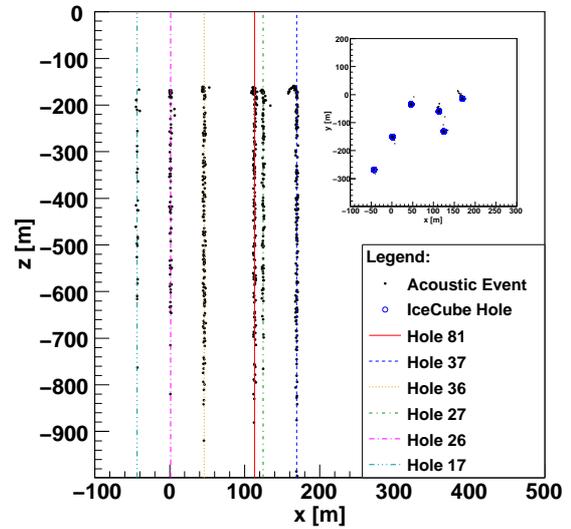}
     \label{fig:recosim}
   \caption{Reconstructed simulated events for visible holes of the drill
 season 08/09}
 \end{figure}
Although knowing that the true IceCube hole diameter is about 70 cm, we take
 into account the possibility that tension cracks might appear outside
 the hole bounding surface which suggests a larger simulation radius.     
The reconstruction of events simulated with constant speed of sound
 and without considering attenuation effects implies an exact source
 localization, which is in contradiction to the real data vertex results, where
 a specific data spread around the source
 (Fig.\ref{fig:all-4s-XY}) and a lack of events below and above a certain depth
 (Fig.\ref{fig:h81_xz_color}) is visible.
The major reason for misreconstruction of events at shallow depth
 (-200 m $<$ z $<$ 0 m) is probably the depth dependence of sound speed 
\cite{SPATSSPEED} which is, therefore, included in the simulation. 
Above 170 m depth, the parametrisation 
$v_s=-(262.379+199.833 ~|z|^{\frac{1}{2}}-1213.08 ~|z|^{\frac{1}{3}})
{\rm m s^{-1}}$
and below a constant sound speed of $v_s=3850~{\rm m s^{-1}}$ is used.
 Further improvement is achieved using additional information
 on sound pressure wave attenuation in the ice \cite{SPATSATT}.
We apply $S_R^n=(S_0 \cdot d_0/d_n)e^{-d_n/\lambda}$
with the initial amplitude $S_0$ at the distance $d_0$ chosen to fit the real
 data \cite{AMPSIM} and an attenuation length $\lambda$ of 300 m as measured
 for South Pole ice with SPATS \cite{SPATSATT}.
If the signal strength at a sensor is above $\sim 300$ mV ($5.2~ \sigma$ above
 the noise level)
 a hit is triggered as in real data. 

A good agreement between reconstructed real and reconstructed simulated events
 is finally obtained as one can see by comparison of Fig.\ref{fig:recosim}
 with Fig.\ref{fig:all-4s-XY} and Fig.\ref{fig:h81_xz_color}.
As expected we observe a large influence by the depth dependent sound speed on
 reconstructions in the upper region of SPATS (between 0 and 200 m depth). 
The significant deviation of the sound speed from the constant value used in
 the reconstruction explains the vertex spreading seen in the real data,
 whereas the direction of these ``smearings'' is caused by the detector
 geometry and points towards the center of SPATS.
The inclusion of the attenuation length makes it more difficult to 
 observe deep events which is in agreement with the real data distribution.

\section{Noise from regions outside IceCube}\label{limits}

In order to determine the number of events not connected to IceCube
 construction activities in the sensitive region of SPATS we omit the area of
 IceCube strings and the data-files from the drill periods keeping in mind that
 we have a lot of acoustic hits here due to detector construction.
Furthermore we look at depths between 200 and 1000 m, in the region of
 constant speed of sound, to avoid the smearing described in section \ref{sim}.
In the 245 days of transient data taking we found no events for the recent 
SPATS-12 detector configuration in the region defined above which allows to
 determine a limit on the cosmogenic neutrino flux.

In order to determine a flux limit an effective volume for the
 SPATS-12 configuration is calculated.
Neutrino interaction vertices were simulated in a cylindrical volume with an
 energy dependent sensitivity radius between $0.4-2~{\rm km}$ around the
 centre of SPATS and at depth between 200 and 1000 m. 
We omit again the area of IceCube strings. 
The neutrinos were assumed to be down going with random $\theta$ and $\phi$. 
Together with the interaction vertex, the direction ($\theta,\phi$) defines the
 plane of acoustic pressure wave. 
The sensor observation angle was then calculated relative to this plane for
 each vertex. 
$10^7$ events were simulated for neutrino energies $E_{\nu}$ increasing from
 $10^{18} {\rm eV}$ to $10^{22} {\rm eV}$. 
The hadronic component of the neutrino energy $E_{had}$ was assumed to be a
 constant fraction $y$ of the neutrino energy i.e. $E_{had}=0.2E_{\nu}$.
In this simulation each acoustic sensor is simply a point at which to calculate
 the acoustic pressure $P_{max}$ as a function of $\theta$ and $R$.
To get a reasonable value for the signal strength at the sensors we apply
 the Askaryan \cite{ASKARYAN} model assuming a cylindrical energy
 deposition in the medium of length L and diameter d.  
No attempt has been made to model angular sensitivity or frequency response. 
A minimum threshold of $\sim$ 300 mV, like in the real
 SPATS-12 measurement is used, which transforms to a necessary minimum pressure of $\sim$ 70 mPa \cite{Karg}.
The observation of zero events inside the effective volume of SPATS-12 gives
 an upper limit of $N_{obs}=2.44$ events by a poissonian 90 \% confidence level \cite{Feldman}.
\begin{figure}[hbtp]
  \centering
    \includegraphics[width=8.6cm]{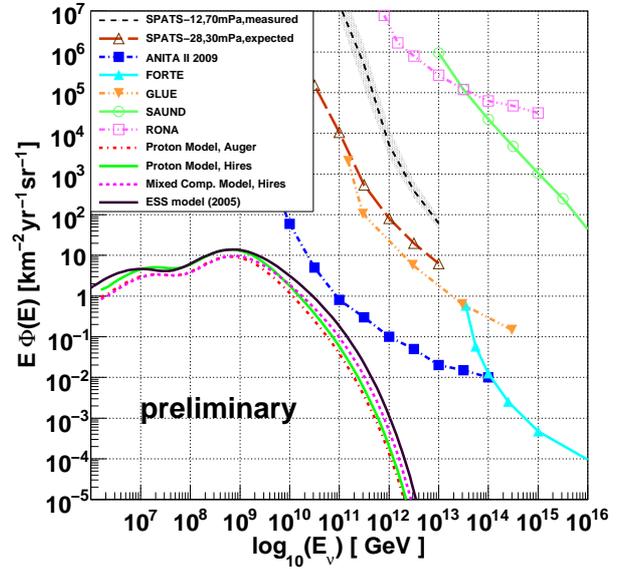}
    \label{fig:sensitivity}
  \caption{Neutrino flux limit of the recent SPATS configuration (70 mPa
 threshold, $\geq 5$ hits per events). 
The gray band (50-100 mPa threshold) around this limit considers uncertainties
 in absolute noise \cite{Karg} and attenuation length \cite{SPATSATT}. }
\end{figure}
The flux limit is shown in Fig.\ref{fig:sensitivity} together with
 the cosmogenic neutrino flux predictions and results of other experiments.
Additionally we show in Fig.\ref{fig:sensitivity} the sensitivity for a
 SPATS-28 detector configuration which potentially would operate
 at a lower threshold of 30 mPa using a new detector readout.








\end{document}